\documentclass[12pt,a4paper]{article}
\usepackage[dvips]{graphicx}
\usepackage{amsmath}
\usepackage{amssymb}
\usepackage{cite}
\newcommand{\rd}{{\rm d}}
\newcommand{\re}{{\rm e}}
\newcommand{\ri}{{\rm i}}
\newcommand{\rB}{{\rm B}}
\newcommand{\dd}{d}

\DeclareGraphicsExtensions{.eps}

\begin{document}
\title{Quantum transport and localization in biased periodic structures under bi-
and polychromatic driving}
\author{A Klumpp\footnote{Present address: Institut f\"ur Physik,
Universit\"at Kassel, D-34109 Kassel, Germany}, D Witthaut, and H J Korsch\\
FB Physik, Technische Universit\"at Kaiserslautern\\ D-67653 Kaiserslautern, Germany}

\maketitle

\begin{abstract}
We consider the dynamics of a quantum particle in a one-di\-men\-sional periodic
potential (lattice) under the action of a static and time-periodic field. The analysis
is based on a nearest-neighbor tight-binding model which allows a convenient closed form
description of the transport properties in terms of generalized Bessel functions. The case
of bichromatic driving is analyzed in detail and the intricate transport and localization
phenomena  depending on the communicability of the two excitation frequencies and the 
Bloch frequency are discussed. The case of polychromatic driving is also discussed, in
particular for flipped static fields, i.e.~rectangular pulses, which can support an
almost dispersionless transport with a velocity independent of the field amplitude.  
\end{abstract}

\section{Introduction}
\label{s-intro}
Quantum transport properties in periodic structures (lattices) are highly non-intuitive,
in particular in view of their localization properties (see, e.g.,\cite{Grif98}). 
Classically, a particle in a
periodic potential is localized in a single well at low energies, whereas a quantum
system shows the well known Bloch band/gap structure with transporting bands. If a constant
field $F_0$ is added, the quantum states in a band with width $G$ become localized
to an interval $G/F_0$, contrary to our intuition. In this region, the wave functions
show the celebrated Bloch oscillations \cite{04bloch1d} with the Bloch
frequency $\omega_\rB=F_0d/\hbar$, where $d$ is the period
of the potential, and there is no directed transport. This localization is, of course, 
approximate because of the decay to
infinity by Zener tunneling \cite{02wsrep}, an effect of longer time scales at least for weak 
fields. 
An additional time-periodic driving $F(t)$ re-introduces the quantum transport
which can be suppressed again for special choice of the parameters.
In the following we will neglect the decay and confine ourselves to the minimal
model system, the single-band tight-binding hamiltonian with  nearest-neighbor coupling
\begin{equation}
\label{Hamtb}
\hat H=-\frac{G}{4} \sum_{\ell=-\infty}^{+\infty} \big(\,
|\ell\rangle \langle \ell +1| +
| \ell +1\rangle \langle \ell|\,\big)
+\dd F \sum_{\ell=-\infty}^{+\infty} \ell |\ell \rangle \langle \ell |\,,
\end{equation}
where $|\ell\rangle$ are the Wannier states which are exponentially
localized in the $\ell$-th potential well and $G$ is the width of the
Bloch band for $F=0$.
Let us recall that for a static field $F=F_0$ the extension of the Bloch oscillation is
approximately
\begin{equation}
\label{Lbloch}
L=G/F_0
\end{equation}
as easily deduced from the tilted band picture.

For the best understood case of a monochromatic driving
$F(t)=F_0-F_1\cos (\omega_1t)$ where the driving frequency
is in resonance with the Bloch frequency, $\omega_\rB=n\omega_1$, 
$n=1,2,\ldots$,
a suppression of transport, denoted as a {\it dynamic localization} is 
found if the condition
 \begin{equation}
\label{dynloc}
J_n(dF_1/\hbar \omega_1)=0
\end{equation}
is satisfied \cite{Dunl86, Holt96,Grif98}, where $J_n(z)$ is the (ordinary) Bessel function.
Otherwise transport with a velocity
proportional to $J_n(dF_1/\hbar \omega)$ is found.

For the general case of a bi- or polychromatic periodic driving the particle
dynamics is found to be quite complicated and unexpected effects have been observed. 
These effects may be used to manipulate the quantum dynamics.
The aim of the present study is to investigate the transport and localization
properties for a bi-chromatic driving extending previous studies
by Liu and Zhu \cite{Liu00}, Suqing {\it et al.} \cite{Suqi03} and
Yashima {\it et al.} \cite{Yash03}. 
In addition, we will briefly consider the even less explored realm of polychromatic
driving

We base our analysis on a Lie-algebraic
approach introduced by one of the authors \cite{03TBalg} (see also \cite{04tb2d}
for an extension to two space dimensions).
Following Ref.~\cite{03TBalg}, we introduce the abbreviations
$f_t=dF(t)/\hbar$ and $g=-G/4\hbar$ which may also depend on time
in general. Then the
Hamiltonian can be conveniently expressed in terms of the hermitian position
operator $\hat N=\sum_\ell \ell\,|\ell\rangle \langle \ell|$ and the unitary shift operator
$\hat K=\sum_\ell |\ell\rangle \langle \ell+1|$:
\begin{equation}
\frac{1}{\hbar}\,\hat H=g\,\big(\hat K+\hat K^\dagger\big)+f_t\,\hat N\,.
\end{equation}
The dynamics of expectation values depends on the initial state
 \begin{equation}
|\psi(t=0)\rangle = \sum_\ell c_\ell |\ell\rangle
\end{equation}
characterized by the coherence parameters
\begin{eqnarray}
K&=&\sum_\ell c_{\ell-1}^*c_\ell=|K|\,\re^{\ri \kappa}\label{cohpar}\\
L&=&\sum_\ell c_{\ell-2}^*c_\ell=|L|\,\re^{\ri \nu}\\
J&=&\sum_\ell (2\ell-1)c_{\ell-1}^*c_\ell=|J|\,\re^{\ri \mu}\,.
\end{eqnarray}
In the following we will assume a symmetric normalized Gaussian state
$c_\ell \sim \re^{-\ell^2/4\sigma^2}$ which allows for a broad initial
distribution ($\sigma \gg 1$) an approximate evaluation of the coherence
parameters by replacing sums by integrals:
\begin{eqnarray}
K\approx\re^{-1/8\sigma^2}\ ,\quad
L\approx \re^{-1/2\sigma^2}\ ,\quad
J\approx 0\,.
\end{eqnarray}
The time evolution operator can be written as a product of exponentials \cite{03TBalg}
\begin{equation}
\hat U(t)=\re^{-\ri \eta_t\hat N}\,
\re^{-\ri\,\chi_t\hat K} \,\re^{-\ri\,\chi_t^*\hat K^\dagger}\,, 
\label{Ut}
\end{equation}
with
\begin{equation}
\eta_t=\int_0^t\! f_\tau \,\rd \tau \quad \textrm{and}\quad
\chi_t=g\int_0^t\! \re^{-\ri \,\eta_\tau}\,\rd \tau 
=|\chi_t|\,\re^{-\ri \phi_t}\,.
\label{eta-chi}
\end{equation}
An advantage of the product form (\ref{Ut}) is the simple evaluation
of matrix elements and expectation values, as for instance the mean value of the 
position $\hat N$,
\begin{eqnarray}
\langle \hat N \rangle_t
=\langle \hat N \rangle_0 \!+\!\ri\,\big(\,\chi_t\,\langle \hat K \rangle_0 \!-\!
\chi_t^*\,\langle \hat K^\dagger \rangle_0\,\big)
=\langle \hat N \rangle_0 \!+\!2|K|\,|\chi_t|\,\sin (\phi_t-\kappa)
\label{Ntav}
\end{eqnarray}
with the coherence parameter $K=|K|\,\re^{\ri \kappa}$ specified in (\ref{cohpar}).
For a symmetric initial state we have $\kappa=0$ and an elementary
calculation provides an upper bound of the mean transport velocity
\begin{equation}
v_{\rm trans}=\frac{\langle \hat N \rangle_t-\langle \hat N \rangle_0}{t}
\le 2|gK|\,.
\end{equation}
The time evolution of the width of the wave packet is given by \cite{03TBalg}
\begin{eqnarray}
\Delta_N^2(t)&=&\Delta_N^2(0)
+2|\chi_t|^2\big\{ 1 -|L|\cos (2\phi_t-\nu)-2|K|^2\sin^2(\phi_t-\kappa)\,\big\}\nonumber\\
&&\qquad +2|\chi_t|\big\{2\langle N\rangle_0|K|\sin (2\phi_t-\kappa)+|J|\sin^2(\phi_t-\mu)\,\big\}\,.
\end{eqnarray}
which simplifies to
\begin{equation}
\Delta_N^2(t)=\Delta_N^2(0)
+2|\chi_t|^2\big\{ 1 -|L|\cos (2\phi_t)-2|K|^2\sin^2(\phi_t)\,\big\}
\end{equation}
for a real symmetric Gaussian initial wave packet. With $|K|=\re^{-1/8\sigma^2}$ and 
$|L|=\re^{-1/2\sigma^2}$ the dispersion depends on $\sigma$ as
$\Delta_N^2(t)-\Delta_N^2(0)\sim 1/\sigma^2$ for large values of $\sigma$.
This leading order term can, however, be
suppressed at times $t$ for which 
%we have $\phi_t=(n+1/2)\pi$, $n\in \mathbb Z$
the function $\chi_t$ is purely imaginary, i.e.
$\cos (2\phi_t)=-1$ and $\sin^2(\phi_t)=1$. One can easily show that this implies
\begin{equation}
\Delta_N^2(t)-\Delta_N^2(0)=
|\chi_t|^2\,\frac{1}{4\sigma^4}+O(\sigma^{-6})
\label{red-disp}
\end{equation}
and the dispersion is strongly reduced at these times. This condition will be satisfied
for the example of the shuttling transport discussed below in section \ref{ss-flip}.
\section{Time periodic driving}
\label{s-periodic}
In the following, we will assume a time-independent nearest-neighbor coupling
$g=g>0$ and a periodic
driving term 
\begin{equation}
f_t=f_0+\tilde f_t\ ,\quad \tilde f_{t+T}=\tilde f_t\,,
\label{f-tilde}
\end{equation}
where $f_0\le 0$ is constant and $\tilde f_t$ vanishes on the average.
This implies a Bloch frequency $\omega_\rB=f_0$. 
 
It can easily be shown by Fourier expansion that both functions in (\ref{eta-chi})
can be separated into a linear growing and a periodically
oscillating part (cf.~\cite{03TBalg}):
\begin{equation}
\eta_t=f_0t+\tilde \eta_t\ ,\quad \chi_t=\gamma t/2+\tilde \chi_t\
\label{eta-chi-tilde}
\end{equation}
where $\gamma$ is different from zero if the driving period 
$T$ is in resonance with the Bloch period $T_\rB=2\pi/\omega_\rB$,
\begin{equation}
T=n\,T_\rB \ ,\quad n\in \mathbb N\,.
\label{res-driving}
\end{equation}
The coefficient $\gamma$ controlling the transport properties is given by 
\begin{equation}
\gamma=\frac{2\chi_T}{T}=\frac{2g}{T}\int_0^T\re^{-\ri n\omega t-\ri \tilde \eta_t}\rd t 
\label{gamma-chi-tilde}
\end{equation}
with $\omega=2\pi /T$. At time $T$ the time evolution operator (\ref{Ut})  commutes with
$\hat K$ and the simultaneous eigenstates of $\hat K$ and $\hat U(T)$ have eigenvalues 
$\re^{\ri \kappa \dd}$ and $\re^{-\ri \epsilon T}$, respectively, where the
quasienergy $\epsilon$ depends on the quasimomentum $\kappa$ as
\begin{equation}
\epsilon(\kappa)=|\gamma | \,\cos (\kappa \dd +\arg (\gamma)\,)
\label{quasienergy}
\end{equation}
denoted as the (quasi)dispersion relation. It may be of interest to realize that the possible values
of $\gamma$ are bounded by the width of the dispersion relation for the non-biased system
\begin{equation}
|\gamma|\le 2|g|\,,
\label{gamma-bound}
\end{equation}
a well known property of Fourier integrals.

The mean value of the wave packet moves with an average velocity 
\begin{eqnarray}
V_{\rm trans} &=& \frac{\langle \hat N \rangle_T d}{T}=  v_{\rm trans}\,d \nonumber \\
v_{\rm trans} &=& \frac{\ri }{T}\,\Big(\,\chi_T\,\langle \hat K \rangle_0 \!-\!
\chi_T^*\,\langle \hat K^\dagger \rangle_0\,\Big)
=-|\gamma|\,\sin \big(\kappa \dd +\arg(\gamma)\,\big)\,,
\label{velocity}
\end{eqnarray}
where
the initial condition is written as $\langle \hat K \rangle_0 =\re^{\ri
\kappa \dd}$.
As expected, this mean velocity can also be expressed as 
$v_{\rm trans}=\rd \epsilon/\rd \kappa|_{\kappa}$\,. 
Note that  $v_{\rm trans}$ is bounded from above
\begin{equation} 
v_{\rm trans}\le |\gamma|\le 2|g|\ \textrm{or}\quad
V_{\rm trans}\le  2d|g|\,.
\label{v-bound}
\end{equation}

If the system parameters are adjusted such that $\gamma =0$,
the band (\ref{quasienergy}) collapses and the non-oscillatory term in 
$\chi_T$ vanishes with the consequence that
there is no transport, i.e.~we have dynamic localization.
\section{Bichromatic driving}
\label{s-bicromatic}
Let us consider a combined dc- and bichromatic ac-driving 
\begin{equation} 
f_t=f_0-f_1\cos (\omega_1t)-f_2\cos (\omega_2t+\delta)
\label{bi-driving}
\end{equation}
with a phase shift  $\delta$.
Note that this function is aperiodic for incommensurable frequencies
$\omega_1$ and $\omega_2$,
which is not necessarily true for a a sum of two arbitrary periodic
functions with incommensurable periods; an example can be found in \cite{Levy05}.
For commensurable frequencies $f_t$ is periodic.

With \,$f_0=\omega_\rB$\, and \, $u=f_1/\omega_1$, $v=f_2/\omega_2$ we obtain 
\begin{eqnarray}
\eta_t=\int\limits_0^t  f_\tau \,\rd\tau
=\omega_\rB t-u\,\sin(\omega_1t)-v\,\sin( \omega_2t +\delta)\,.
\end{eqnarray}
The function $\chi_t$ in eq. (\ref{eta-chi}) which determines
the dynamics can be expressed in terms of Bessel functions using the generating function
\begin{eqnarray}
\re^{\ri x\sin{z}}=\sum\limits_{\mu=-\infty}^\infty 
J_\mu\left(x\right)\re^{\ri \mu z}\,.
\end{eqnarray}
With the abbreviation \,$\omega_{\mu ,\nu}=\omega_\rB-\mu\omega_1-\nu\omega_2$\,
this yields
\begin{eqnarray}
\nonumber
\chi_t&=&\int\limits_0^t g\re^ {-\ri \eta_\tau}\,\rd\tau
=g\int\limits_0^t \re^{-\ri(\omega_\rB t-u \sin(\omega_1t)-
v\sin(\omega_2t +\delta)\, )}\,\rd \tau\\
&=&g\sum\limits_{\mu =-\infty }^\infty \sum\limits_{\nu =-\infty}^\infty 
J_\mu (u)J_\nu (v)\,\re^{\ri \nu\delta}\int\limits_0^t d\tau \,\re^{-\ri\omega_{\mu ,\nu
}\tau} \,.\label{chibi}
\end{eqnarray}
The remaining integration is elementary, however one has to distinguish 
the cases of resonant and non-resonant driving. 

In the resonant case, we have
\begin{equation}
\omega_\rB=\mu\omega_1+\nu\omega_2 \label{Reso}
\end{equation}
for certain integers $\mu ,\nu$. Let us denote the set of all indices $\mu ,\nu$
satisfying (\ref{Reso}) by $\mathbb{B}$.
We then have
\begin{eqnarray}
\chi_t=\gamma t/2+2g\sum\limits_{(\mu,\nu)\notin\mathbb{B}} 
J_\mu(u)J_\nu(v)\,\re^{\ri \nu\delta}\,\frac{1}{\omega_{\mu,\nu}}
\,\re^{-\ri\omega_{\mu,\nu}t/2}\sin{\frac{\omega_{\mu,\nu}t}{2}},\label{chizerl}
\end{eqnarray}
with
\begin{eqnarray}\label{Gammf1}
\gamma =\frac{2\chi_T}{T}=2g \sum\limits_{(\mu,\nu) \in \mathbb{B}} 
J_\mu (u)J_\nu (v)\,\re^{\ri \nu\delta}\,.
\end{eqnarray}
If $\mathbb B\neq \emptyset$ the first term in (\ref{chizerl}) is growing linearly in
time and thus dominating the long time dynamics. In this case we can observe transport.

It is instructive to have a brief look at the monochromatic driving first. This case
is recovered for $v=0$, observing $J_\nu(0)=\delta_{\nu,0}$ and hence
\begin{eqnarray}
\chi_t=\gamma t/2+2g\sum\limits_{\mu \notin\mathbb{B}}
J_\mu(u)\frac{1}{\omega_{\mu}}\,\re^{-\ri\omega_{\mu}t/2}\sin{\frac{\omega_{\mu}t}{2}},\label{chizerlmono}
\end{eqnarray}
with $\omega_\mu=\omega_\rB-\mu\omega_1$ (cf.~\cite{Grif98,03TBalg})
If the resonance condition $\omega_\rB=\mu \omega_1$ is satisfied, we have
\begin{eqnarray}
\label{Gammf2}
\gamma=2g J_\mu (u)
\end{eqnarray}
and therefore an average transport with velocity proportional to the Bessel function
\,$J_\mu(u)=J_\mu(f_1/\omega_1)$. 
This linear transport is superimposed by an oscillating part, which is periodic with the
Bloch period $T_\rB$ for resonant driving.
By adjusting
the value of the driving amplitude $f_1$ one can change the direction of the transport or bring it
to a standstill for $J_\mu (f_1/\omega_1)=0$, the dynamic localization. 

For bichromatic driving, the situation is more complicated. We distinguish
two cases:\\[2ex]
{\it Incommensurable driving frequencies $\omega_1$ and $\omega_2$:}\\[1mm] 
If the ratio 
$\omega_2/\omega_1$ is irrational,
the resonance condition (\ref{Reso}) leads to 
$\omega_\rB/\omega_1=\mu +\nu \omega_2/\omega_1$. This can only be satisfied for 
a special irrational value of the
ratio $\omega_\rB/\omega_1$, for example by an appropriate dc-field $F_0$.
These cases are of measure zero and {\it typically} there is {\it no transport}.
If, however, the system parameters are tuned to special values satisfying the resonance 
condition (\ref{Reso}), then
there exists only a single  pair $(\mu,\nu)\in\mathbb B$  as can be easily checked and
we have transport with velocity
\begin{eqnarray}\label{Gammf3}
\gamma =2g J_\mu (u)J_\nu (v)\,\re^{\ri \nu\delta}\,.
\end{eqnarray}
\noindent
{\it Commensurable driving frequencies $\omega_1$ and $\omega_2$:}\\[1mm] 
If $\omega_2/\omega_1=q/p$ for integers $q$ and $p$ with
no common divisor and if the resonance condition (\ref{Reso}) is satisfied for a particular pair 
$(M,N)$ one can find all solutions by \,$(\mu,\nu)=(M-qk,N+pk)$\, with \,$k \in \mathbb{Z}$, i.e.
\begin{equation}
\mathbb B = \left\lbrace (M-qk,N+pk), k \in \mathbb{Z} \right\rbrace .
\end{equation}
This can be shown as follows:
The resonance condition (\ref{Reso}) can be rewritten as the Diophantine equation
\begin{equation}
\label{ganzrat}
n=p\mu +q\nu \quad\mbox{with}\quad n=p\,\omega_\rB/\omega_1\in \mathbb{Z}\,.
\end{equation}
A solution  $(\mu,\nu)=(M,N)$ of this equation always exists if $p$ and $q$ have no common divisor
and can be found systematically by, e.g., the Euclid algorithm \cite{Mord69}.
We therefore have an infinite number of solutions.
Note that the choice of the special solution $(M,N)$ is arbitrary. 
For rational frequency ratios the function (\ref{Gammf1}) is determined
by three integers, $p$, $q$, $n$, and we can express it in the convenient form
\begin{eqnarray}\label{Besselfkt}
\gamma &=& 2 g\, \re^{\ri N\delta}\,J_{n}^{p,q}(u,v;\,\re^{\ri p\delta})
\end{eqnarray}
in terms of the two-dimensional, one-parameter Bessel function
\begin{eqnarray}
J_{n}^{p,q}(u,v;z)=\sum\limits_{k=-\infty}^\infty J_{M-qk}(u) J_{N+pk}(v)\,z^k
\label{verallgB}
\end{eqnarray}
where $(M,N)$ is an arbitrary solution of the Diophantine equation (\ref{ganzrat}).
An introduction into the properties of two-dimensional Bessel functions can be found in
\cite{05bessel2d}. 

The two-dimensional generalized Bessel functions (\ref{verallgB}) satisfy the inequality 
\,$|J_{n}^{p,q}(u,v;z)|\le 1/\sqrt{2}$\, for $n\ne 0$. This implies that the maximum
transport velocity
\begin{eqnarray}
v_{\rm trans,max}=|\gamma|=2|g|\,\big| J_{n}^{p,q}(u,v;\re^{\ri p\delta})\big|
\label{vtrans-bicro}
\end{eqnarray}
is bounded as
\,$|\gamma|\le \sqrt{2}g <2g$\,.

For commensurable driving frequencies $\omega_1$ and $\omega_2$,
$\omega_2/\omega_1=q/p$, and driving amplitudes $f_1$ $f_2$ we
have the following behavior:
\begin{itemize}
\item[]
If the ratio $\omega_\rB/\omega_1$ is irrational,
the condition (\ref{Reso}) cannot be satisfied and we have no transport.
\item[]
If $\omega_\rB/\omega_1$ is rational with $n=p\omega_\rB/\omega_1$ we typically have transport
with transport velocity proportional to $J_n^{p,q}(f_1/\omega_1,f_2/\omega_2)$.
This transport can, however,
be stopped for special values of the system parameters and dynamic
localization is observed. It should be noted that this case is always met
for pure ac-driving ($\omega_\rB=0$).
\end{itemize}
The transport and localization properties of a bichromatically driven system are
summarized in Table \ref{table1}. The following section considers some
exemplary cases in more detail.

\begin{table}
\begin{center}
\begin{tabular}{c|c|c}
\rule[-3mm]{0mm}{8mm} & $\omega_\rB/\omega_1$ irrational & $\omega_\rB/\omega_1$ rational\\ \hline
 $\omega_2/\omega_1$ & localization & \\
 irrational & (transport possible) & \raisebox{1.5ex}[-1.5ex]{localization}\\\hline
  $\omega_2/\omega_1$ &  &  transport\\
rational & \raisebox{1.5ex}[-1.5ex]{localization} & (localization possible) 
\end{tabular}
\caption{\label{table1} Transport properties for a bichromatically driven
biased lattice with Bloch frequency $\omega_\rB$.}
\end{center}
\end{table}

\section{Case studies for resonant driving}
\label{s-case}
In this section, we will discuss the transport and localization properties
in some more detail
for the most interesting case of resonant driving
\,$\omega_\rB=\mu\omega_1+\nu\omega_2$\,. The driving frequencies are assumed to be
commensurable,  $\omega_2/\omega_1=p/q$ 
with coprime integers $p$ and $q$, where,  w.l.o.g., we can choose $p$ as
the smaller one of these numbers. Let us recall that the ac-driving period $T$ is
equal to an integer number of Bloch periods, $T=nT_\rB$, where $n$ can be expressed as
$n=pM+qN$ with integer $(M,N)\in \mathbb B$. We will also assume that the force term
is symmetric in time, $f(-t)=f(t)$, i.e.~the phase shift  $\delta$
in (\ref{bi-driving}) is equal to zero, with the consequence that the 
two-dimensional one-parameter
Bessel functions reduce to simple two-dimensional ones \cite{05bessel2d}. The transport 
properties of the system are then determined by the parameter
\begin{eqnarray}\label{Gammaf3}
\gamma &=& 2 g\,J_{n}^{p,q}(u,v)\,.
\end{eqnarray}
with variables $u=f_1/\omega_1$ and $v=f_2/\omega_2$. 
In this case $\gamma$ is real and its sign, the sign of the Bessel function, directly
gives the direction of transport. 
Here we are in particular interested in the
parameter values leading to dynamic localization. These values are given by the zeros
of $J_n^{p,q}(u,v)$, i.e.~the nodal lines of the two-dimensional Bessel
function.

\begin{figure}[t]
\centering
\includegraphics[height=3.8cm,  angle=0]{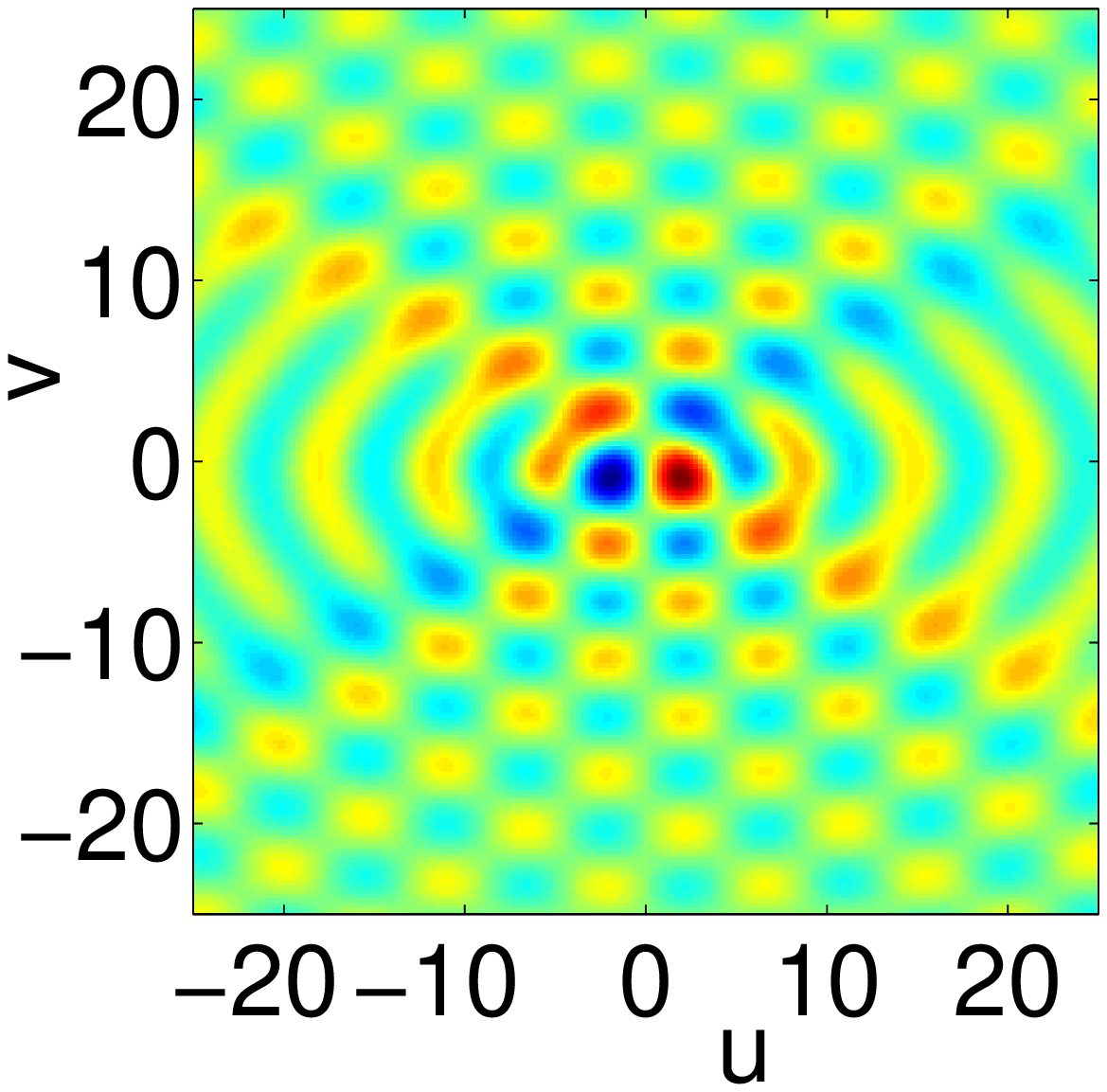}
\hspace{3mm}
\includegraphics[height=3.8cm,  angle=0]{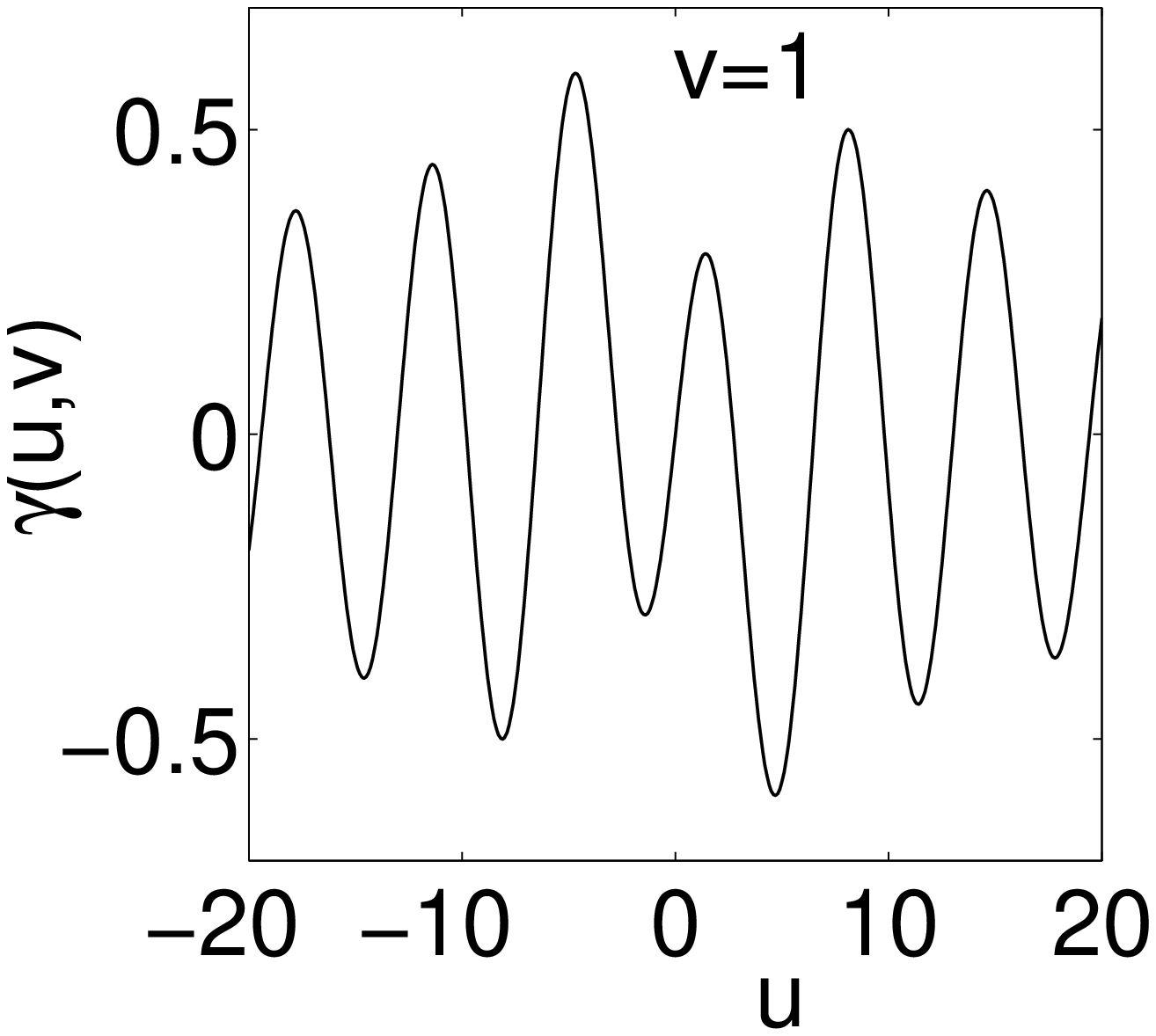}
\hspace{1mm}
\includegraphics[height=3.8cm,  angle=0]{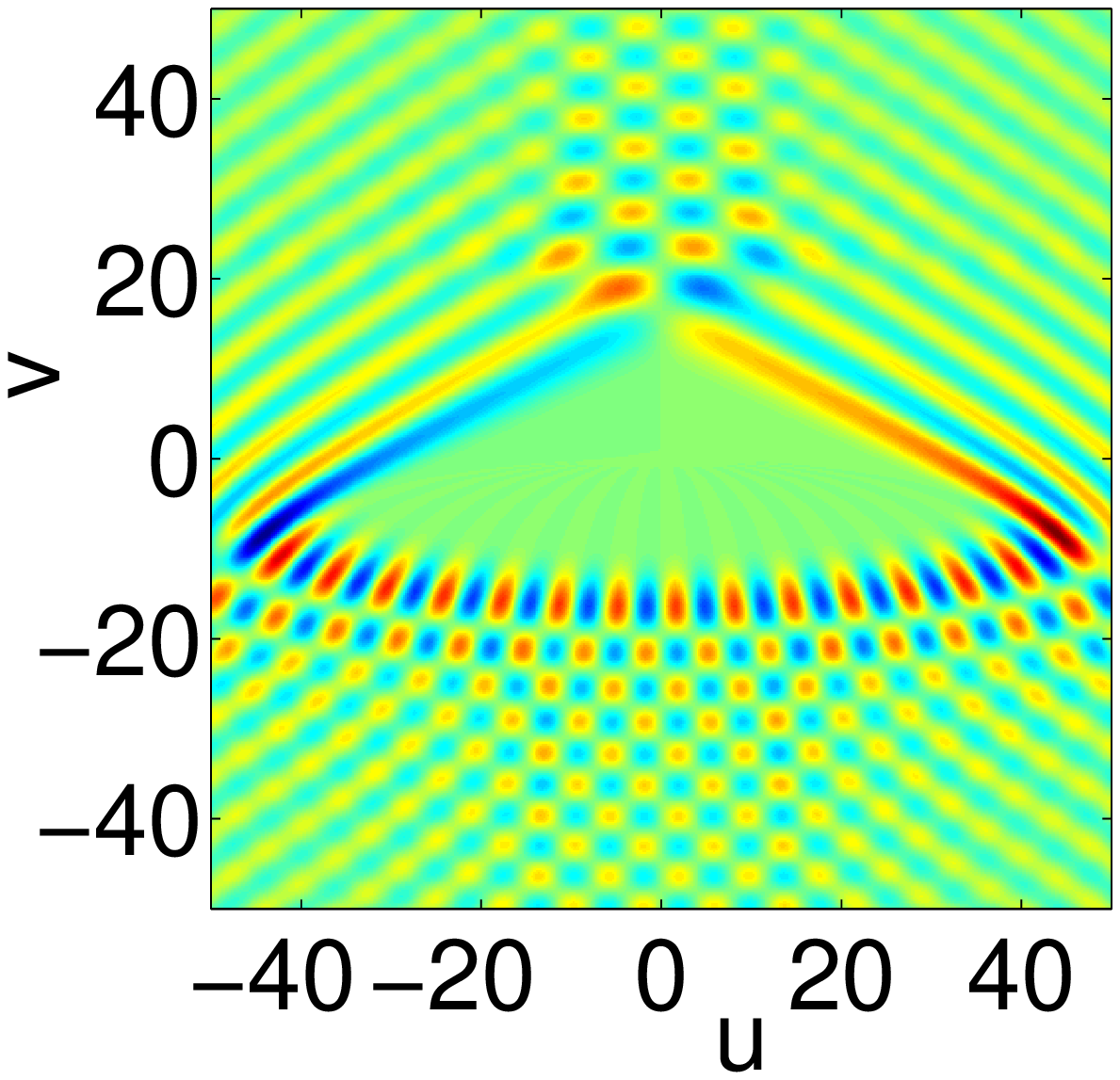}
\caption{\label{fig-bessels}
Transport coefficient $\gamma$ as a function of the parameters
$u =f_1/\omega_1$ and $v=f_2/\omega_2$ for a bichromatic driving
with $\omega_2=2\omega_1$ at resonance with the Bloch frequency
$\omega_\rB$, i.e. \,$\omega_1+\omega_2=n\omega_\rB$.
Shown are examples for $n=1$ (left and middle) and $n=29$ (right),
where the figure in the middle shows $\gamma$ for a fixed parameter
$v=1$.}
\end{figure}

In previous studies the special case of the bichromatic driving with
$\omega_2=2\omega_1$ attracted most attention and we will start with
this special case in the following. For simplicity we fix $g=1$ and
$\omega_1=1$. The resonance condition for the static field
(\ref{ganzrat}) then reads $n =\omega_\rB/\omega_1 = f_0/\omega_1$.
In our first examples we fix the amplitude of one driving as
$f_2 = \omega_2$, i.e $v=1$, and consider the case $\omega_\rB = \omega_1$,
i.e. $n=1$.
The transport coefficient given by $\gamma = 2 J_1^{1,2}(u,v)$ is
shown in figure \ref{fig-bessels} in dependence of $u$ and $v$
(left-hand side) and for $v=1$ fixed (middle). Numerically we find
a global maximum at $u = -4.68$ and a nodal line at $u = -6.49$
for $v=1$.

\begin{figure}[t]
\centering
\includegraphics[width=6cm,  angle=0]{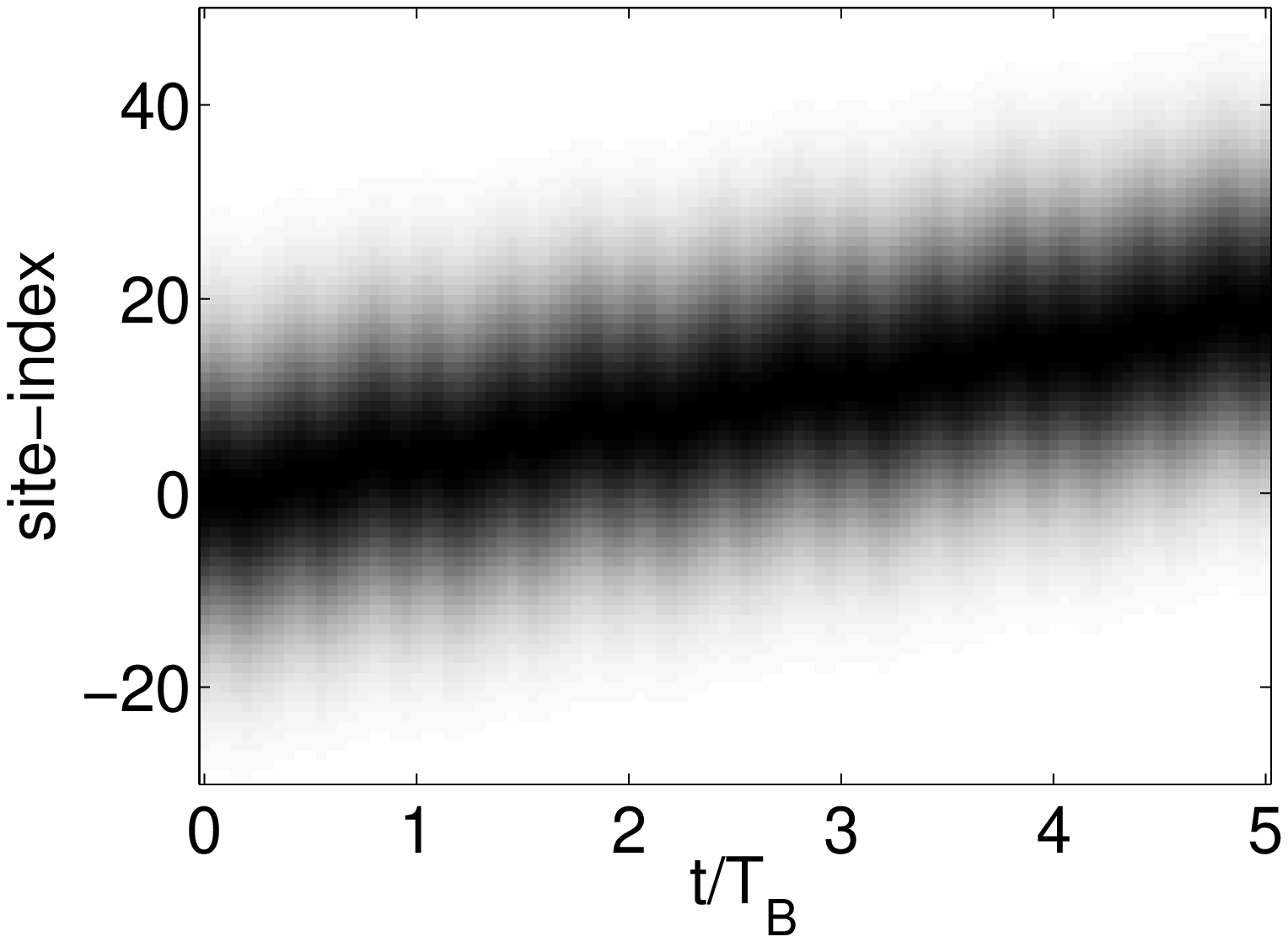}
\includegraphics[width=6cm,  angle=0]{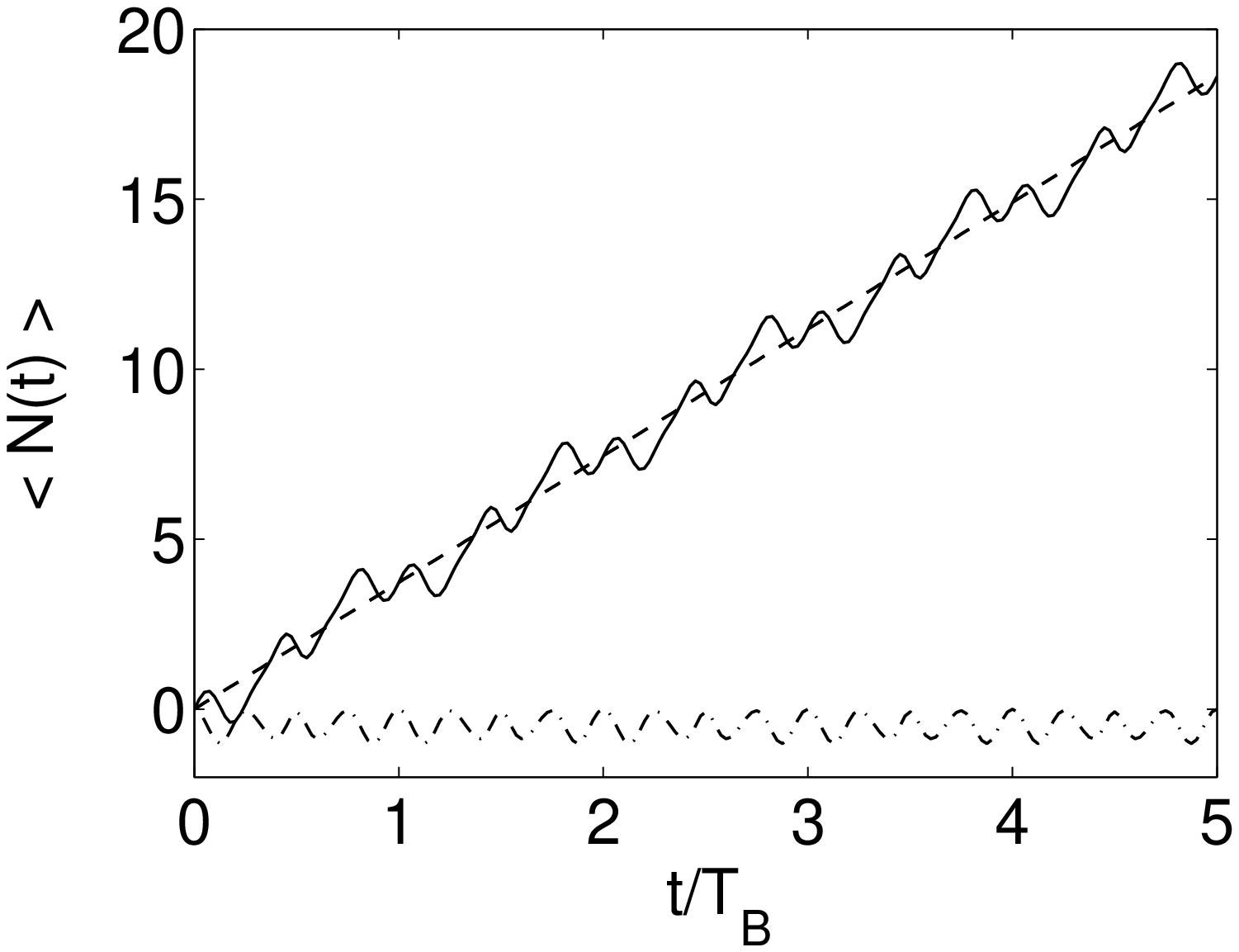}
\caption{\label{fig-n1-maxt}
Left: Dynamics of a gaussian wave packet with initial momentum
$\kappa_0 = -\pi/2$ for a bichromatic driving with amplitudes
$v=1$ and $u = -4.68$ in a greyscale plot.
Right: Position expectation value $\langle \hat N(t) \rangle$
for $\kappa_0 = -\pi/2$ (solid line) and $\kappa_0 = 0$
(dash-dotted line) in comparison with the transport coefficient
$\gamma t$ (dashed line).}
\end{figure}

We consider the time evolution of a broad Gaussian wavepacket
\begin{equation}
  | \psi(t=0) \rangle \sim \sum_\ell \exp\left(-\ell^2/(2\sigma)^2 + \ri \kappa_0 \ell\right)
  | \ell \rangle
\end{equation}
with $\sigma = 10$. For such a broad gaussian wavepacket the coherence
parameter $K$ is given by $|K| \approx 1$ and $\kappa \approx \kappa_0$.
The left-hand side of figure \ref{fig-n1-maxt} shows the dynamics of
such a wave packet for $\kappa_0 = -\pi/2$, which is rapidly transported.
The right-hand side shows the expectation values $\langle \hat N(t) \rangle$
of the position for $\kappa_0 = 0$ and $\kappa_0 = -\pi/2$.
The transport coefficient $\gamma t$ is shown as a dashed line for
comparison.
The sine in equation (\ref{Ntav}) is approximately one for $\kappa_0 = -\pi/2$
such that the fastest possible transport is observed which is given by
$\gamma t$. For $\kappa_0 = 0$ the sine in equation (\ref{Ntav}) is
approximately zero so that the wavepacket stays at rest.
However, this localization is only due to the special initial state
chosen in this example. Depending on the value of $\kappa_0$, the wavepacket
will be transported with a velocity in the interval $[-\gamma,\gamma]$.
Localization, i.e. $v_{\rm trans} = 0$ is only found for $\kappa_0 = 0$.

\begin{figure}[htb]
\centering
\includegraphics[width=6cm,  angle=0]{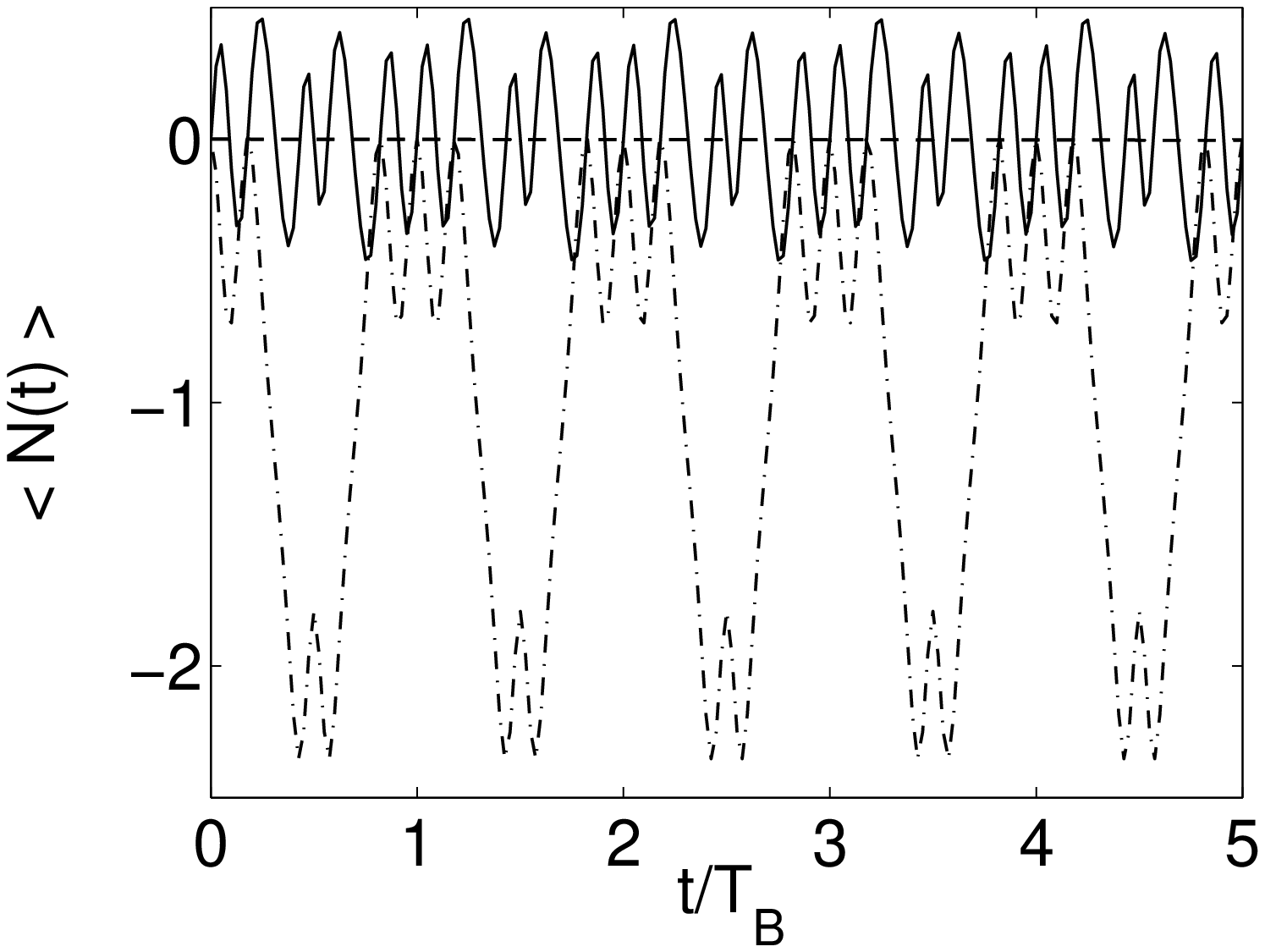}
\includegraphics[width=6cm,  angle=0]{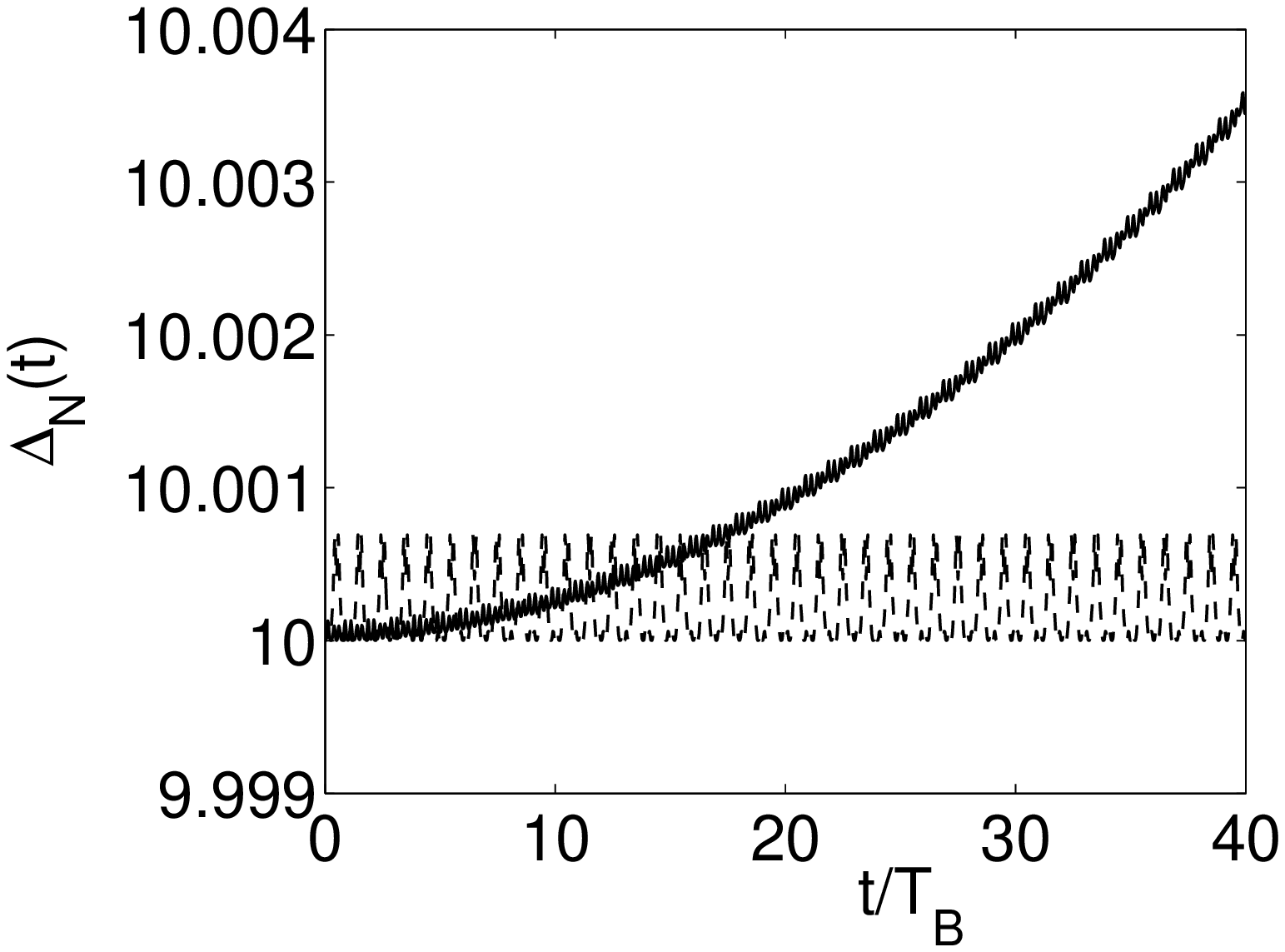}
\caption{\label{fig-n1-dynlok}
Left: Dynamical Localization. For $u = -6.49$ any wave packet is
localized around $n=0$ regardless of the initial momentum.
The evolution of the position expectation $\langle \hat N(t) \rangle$
is shown for $\kappa_0 = -\pi/2$ (solid line) and $\kappa_0 = 0$
(dash-dotted line).
Right: Dispersion of a gaussian wave packet with $\kappa_0 = - \pi/2$
for $u = -4.68$ (transport, solid line) and $u = -6.49$ (dynamical
localization, dashed).}
\end{figure}

The situation is different for $u = -6.49$, for which $J_1^{1,2}(u,v)=0$,
i.e. dynamical localization is found.
Now {\it every} wavepacket will be localized regardless of its initial momentum
$\kappa$. This is illustrated in figure \ref{fig-n1-dynlok} on the left
where the position expectation value $\langle \hat N(t) \rangle$ is shown
for $\kappa_0 = 0$ and $\kappa_0 = -\pi/2$.
The right hand side of figure \ref{fig-n1-dynlok} shows the time evolution
of the width of the wavepacket with $\kappa_0 = 0$ for $u = -4.68$ (transport)
and $u = -6.49$ (dynamical localization). As the dispersion is also
governed by the function $|\chi_t|$, no broadening is observed in
the case of dynamical localization.

\begin{figure}[t]
\centering
\includegraphics[width=6cm,  angle=0]{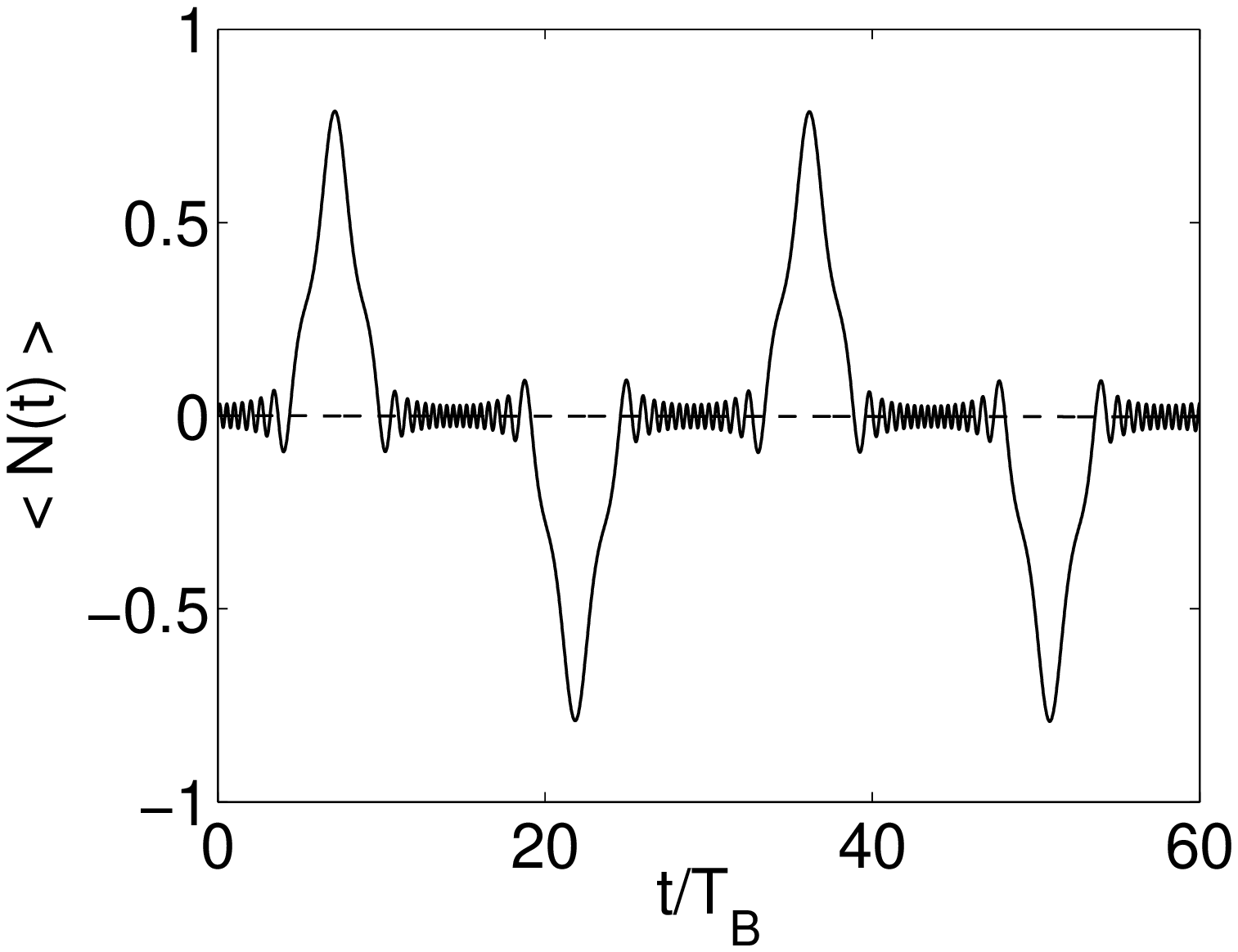}
\includegraphics[width=6cm,  angle=0]{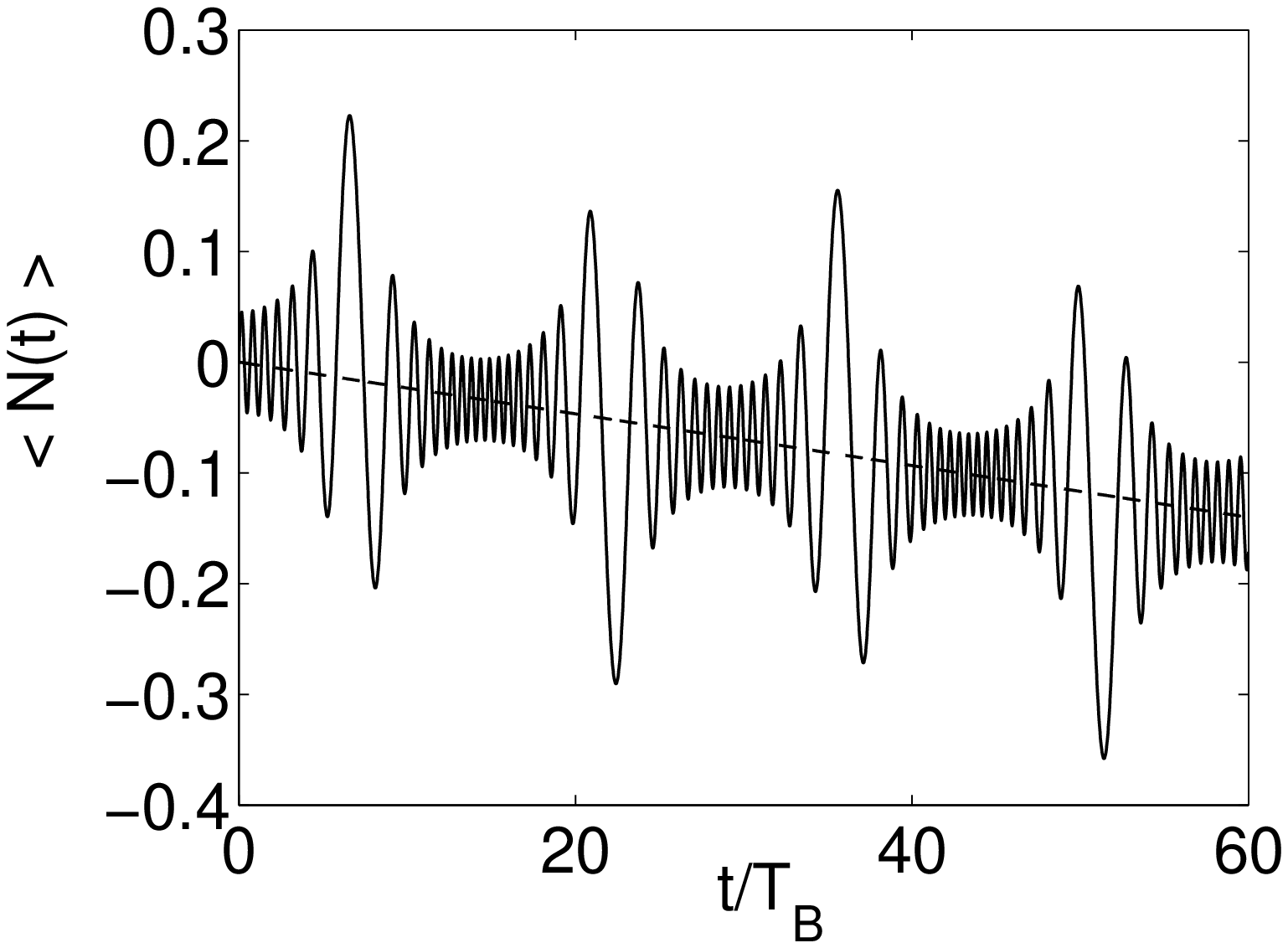}
\caption{\label{fig-n29-lt}
Dynamical Localization for $n=29$, $v=-20$ and $u = 3.37$ (left)
and transport for $u = 5$ (right). The mean transport is given by
$\gamma t$, which is plotted as dashed line for comparison.}
\end{figure}

Next we consider an example, where the field amplitudes are large in
comparison to the driving frequencies. Then $n = f_0/\omega \ll 1$
as well as $u,v \ll 1$ and asymptotic approximations for the relevant
Bessel functions are available \cite{05bessel2d}. In fact we consider
a driving amplitude $v = -20$ and a static field strength with $n = 29$.
Dynamical localization is observed if the driving amplitudes $u$
is chosen such that $J^{1,2}_29(u,v) = 0$. Explicit estimates for
these values are derived from the asymptotic approximations of this
Bessel function. For $n<2|v|$ one finds
\begin{eqnarray}
u\sqrt{\frac12-\frac{n}{4v}}=\left\{
\begin{array}{ll}
(2j+1)\frac{\pi}{2} \quad & n \textrm{   even}\\
j\pi \quad & n \textrm{   odd}\end{array}\right.
\quad ,\quad j=0,\,\pm 1,\,\pm 2,\ldots
\label{zero1}
\end{eqnarray}
and for $n>2|v|$ with $v<0$
\begin{eqnarray}
u\,\sqrt{\frac12-\frac{n}{4v}}=\big( n+2j\big)\frac{\pi}{2}\quad ,\quad j=0,\,\pm 1,\,\pm 2,\ldots\,.
\label{zero2}
\end{eqnarray}
The smallest non-zero values of $u$ for which dynamical localization
is predicted by these formula predict are $u_1 = 3.38$ and $u_2 = 6.77$,
while we find $u_1 = 3.37$ and $u_2 = 6.75$ numerically.
Figure \ref{fig-n29-lt} shows the time evolution of the position expectation
value for $u_1 = 3.37$ (left) and for $u_3 = 5$ (right), each for $\kappa_0 = -\pi/2$.
Clearly the wave packet stays and rest and does not show any systematic
dispersion for $u_1 = 3.37$. Transport is found for $u_3 = 5$, where
the mean transport is given by $\gamma t$.
However, transport is much slower than for $n=1$ as illustrated in
figure \ref{fig-n1-maxt}.

\section{Polychromatic driving}
\label{s-poly}
In view of the rich transport and localization behavior for bichromatic driving, one may
expect additional surprising effects for general time periodic, however polychromatic driving.
This is indeed
the case, however, only few such studies have been reported up to now
in particular for a periodic rectangular force \cite{Zhu99,04bloch1d}. This case
allowing a closed form solution, is discussed in the following section. The general
case is briefly outlined in sect.~\ref{ss-fourier}.
\subsection{Flipped fields and shuttling transport}
\label{ss-flip}
Let us consider a rectangular force term
\begin{eqnarray}
f(t)=\left\{\begin{array}{ll}
 f_1 \qquad & 0\le t< aT\\[1ex]
f_2 \qquad & aT\le t<T\\[1ex]
 \end{array}\right.\label{ft-rec}
\end{eqnarray}
with $0 < a < 1$ and $f(t+T)=f(t)$. For this flipped static force
excitation the time-averaged field is
\begin{eqnarray}
f_0=af_1+bf_2 \quad ,\quad b=1-a\,,
\end{eqnarray}
where we can assume $f_1>0$. 
For the choice $f_2=-f_1$, the force (\ref{ft-rec}) reduces to
the case studied in \cite{Zhu99} (note that there the time scale
is shifted). The even more specialized case $a=1/2$ with period
$T=2\pi/f_1$, the Bloch period for a constant field $f_1$, is discussed
in \cite{Lenz03} and briefly at the end of \cite{04bloch1d} (note that in this 
case we have $f_0=0$).  

For the flipped field excitation (\ref{ft-rec}), the functions 
$\eta_t$ and $\chi_t$ in (\ref{eta-chi})
can be calculated by elementary integration: 
\begin{eqnarray}
&&\eta_t=\left\{\begin{array}{ll}
f_1t \qquad & 0\le t\le aT\\[1ex]
f_1 aT+ f_2(t-aT)\qquad & aT\le t\le T\\[1ex]
 \end{array}\right.\\[2mm]
&&\eta_{t+T}=\eta_t+\eta_T\quad ,\ \eta_T=f_1 aT+ f_2bT=f_0t\,.
\label{etat-flip}
\end{eqnarray}
\begin{eqnarray}
\chi_t=\left\{\begin{array}{ll}\displaystyle
 \frac{g}{\ri f_1}\,\Big(1- \re^{-\ri f_1t}\Big)   \quad & 0\le t\le aT\\[3ex]
 \displaystyle \frac{g}{\ri f_1}\,\Big(1- \re^{-\ri f_1aT}\Big)  
 + \frac{ g\re^{-\ri f_1aT}}{\ri f_2}\,\Big(1- \re^{-\ri f_2(t-aT)}\Big) \quad & aT\le t\le T\\[1ex]
 \end{array}\right.
\label{chit-flip}
\end{eqnarray}
and 
\begin{eqnarray}
&& \chi_{t+T}=\re^{-\ri \eta_T}\chi_t + \chi_T
\label{chiTt-flip}\\[2mm]
&&\chi_T=\frac{g}{\ri f_1}\,\Big(1-\re^{-\ri f_1aT}\Big)
+\frac{g}{\ri f_2}\,\Big(\re^{-\ri f_1aT}-\re^{-\ri f_0T}\Big)\,.
\label{chiT-flip}
\end{eqnarray}
With $f_0=\omega_\rB$ and $\omega=2\pi/T$, we see that  
for the case of resonant driving, $\omega_\rB =n\omega$, we have $f_0T=2\pi n$\,, 
$\chi_{kT}=k\chi_T$ for $k=1,\,2,\,\ldots$. This implies an overall transport
of the expectation value $\langle \hat N \rangle_t$ with a
velocity determined by 
\begin{eqnarray}
\gamma=\frac{2\chi_T}{T}=\frac{4g}{T}\Big(
\frac{1}{f_1} - \frac{1}{f_2}\Big)\ \re^{-\ri \frac{af_1T}{2}}
\,\sin \frac{af_1T}{2}\,.
\label{gammares}
\end{eqnarray}

For $af_1T=2\mu \pi$ with $\mu=1,\,2,\,\ldots$ we have $\gamma=0$, i.e.~dynamical localization.
The maximum transport velocity 
\begin{eqnarray}
v_{\rm trans}=|\gamma|=\Big|\frac{4g}{T}\,
\Big( \frac{1}{f_1} - \frac{1}{f_2}\Big)\,\sin \frac{af_1T}{2}\Big|\le 2|g|
\label{vtrans}
\end{eqnarray}
is obtained when the phase is adjusted as 
$\kappa d+\arg(\gamma)=\kappa d-af_1T/2=(2\mu+1)\pi/2$ according to (\ref{velocity}).

\begin{figure}[t]
\begin{center}
\includegraphics[width=.6\textwidth]{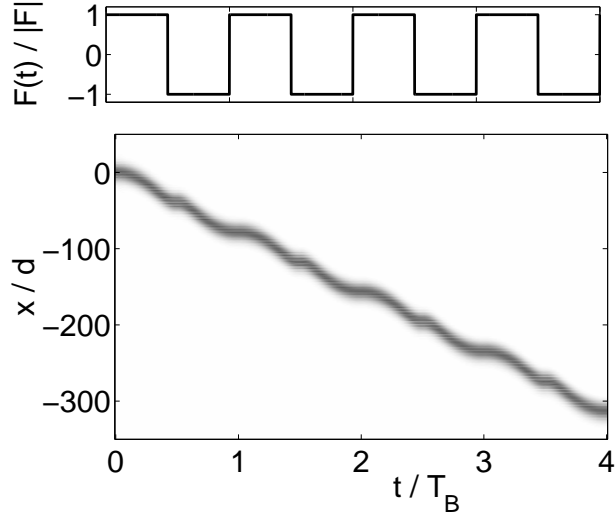}
\end{center}
\caption{\label{flipp1} Numerical time propagation of a Gaussian wave packet
in a cos-potential for a flipped field $f_1=-f_2=0.0003$ and $a=0.5$.}
\end{figure}

These findings can be
easily explained in terms of Bloch oscillations for the time intervals with constant
force $f_1$ and $f_2$ with Bloch periods $T_{\rB 1}=2\pi/f_1$ and $T_{\rB 2}=2\pi/|f_2|$,
respectively. 
If $af_1T=2\mu \pi$ with $\mu\in \mathbb N$, we have $b|f_2|T=2\nu \pi$ with $\nu\in \mathbb N$
for resonant driving, $f_0=af_1T+bf_2T=2n\pi$. The wave packet carries out $\mu$ full
Bloch oscillations in the first time interval and $\nu$ in the second and there is no
no directed transport. For $af_1T=(2\mu+1)\pi$, however, we also have 
$a|f_2|T=(2\nu+1)\pi$ ($\mu,\nu \in \mathbb N$) and therefore a motion over a distance
$L_1=\Delta/F_1=4gd/f_1$ in the first time interval $aT$ and $L_2=\Delta/F_2=4gd/f_2$ in
second time interval $bT$, however in the opposite direction.
The average transport velocity is 
\begin{eqnarray}
V_{\rm trans} =\frac{L_1+L_2}{T}= \frac{4gd}{T}\,\Big|\frac{1}{f_1}-\frac{1}{f_2}\Big|
\end{eqnarray}
in agreement with (\ref{vtrans}).

Let us briefly consider the most simple case where a static field $f$ with
Bloch period $T_\rB=2\pi/f$ is flipped after each half of a Bloch period, i.e.~we
have $f_1=-f_2=f$, $a=1/2$, $f_0=0$ and a period $T=T_\rB$. This yields a transport
velocity
\begin{eqnarray}
v_{\rm trans} =\frac{4|g|}{\pi} ,
\end{eqnarray}
which is by a factor of $2/\pi$ smaller than the maximum possible
value $2|g|$. It is remarkable that the velocity of this field induced
transport is {\it independent\/} of the field amplitude $f$.

\begin{figure}[t]
\begin{center}
\includegraphics[width=.6\textwidth]{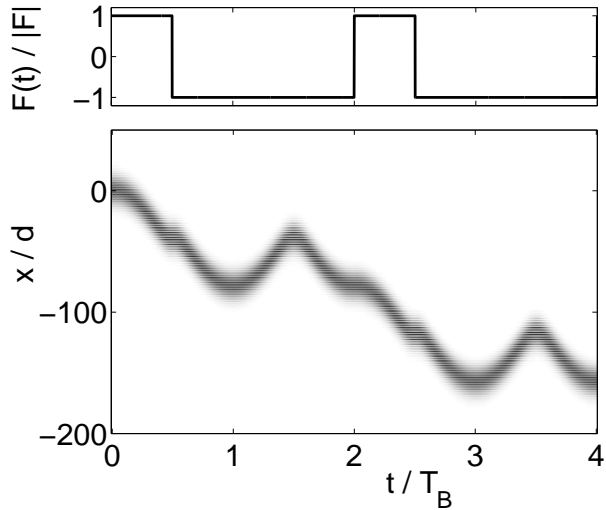}
\end{center}
\caption{\label{flipp2} Numerical time propagation of a Gaussian wave packet for
a flipped field $f_1=-f_2=0.0003$ and $a=1/4$.}
\end{figure}

The analysis above is based on a tight-binding model, Numerically exact
wave packet propagation can be used to check the validity of these predictions
for realistic potentials.
As an illustrating example, figs.~\ref{flipp1} and \ref{flipp2} show the propagation of an
initially broad Gaussian wave packet in position space,
\begin{eqnarray}
\psi(x,t=0) \sim \exp(-x^2/(2s)^2)\ ,\ s = 15\pi\,
\end{eqnarray}
(we use units with $\hbar=1$)
in a potential \,$V(x) = (1/8)\, \cos x$\, for  a flipped $F(t)$ field with
$|F|  = 0.0003$ and $f_1=-f_2$. In the first case shown in  fig.~\ref{flipp1}
the static force is flipped after each half-period of the Bloch time ($a=1/2$),
i.e.~the average field $f_0$ vanishes (see also \cite{04bloch1d} for a similar
calculation). One observes that the wave packet moves almost dispersionless
with a velocity $V_{\rm trans} = -78 \, d/T_B$, which agrees with the
prediction of the tight-binding model for the actual bandwidth
$\Delta = 0.0741$. Note that the direction of transport is determined by the
direction of the Bloch oscillation at time $t=0$, i.e.~ in the negative direction
in the present case with initially positive $f$. In order to demonstrate this clearly,
fig.~\ref{flipp2} shows the same system as before, however for a doubled
excitation period $T$ and $a=1/4$, i.e.~a negative average force $f_0$. 
The wave packet moves in the same
direction as before, i.e.~opposite to the average gradient of the potential.
Note that the transport velocity is reduced by a factor of 2, because of the 
increased value of $T$. This behavior can be easily understood by observing that
in the second half of the excitation period, the system undergoes a full Bloch
oscillation without a net transport.
\subsection{General polychromatic driving}
\label{ss-fourier}
The two simple cases analyzed above, mono- or bichromatic fields
or binary-flipped piecewise constant fields allow a quite extensive 
analytical analysis. Let us finally briefly address a general polychromatic
driving field following \cite{03TBalg}.
For simplicity we will assume $f(t)=f(t+T)$ to be symmetric in time, $f(-t)=f(t)$, with Fourier
expansion
\begin{equation} 
f_t=f_0-\sum_{m=1}^\infty f_m\cos (m\omega t) \ ,\quad \omega=2\pi/T\,.
\end{equation}
Let us recall that in the units chosen here the dc-component is equal to the 
Bloch frequency  $f_0=\omega_\rB$, and the function $\eta_t$ in eq.~(\ref{eta-chi})
is
\begin{equation} 
\eta_t=\omega_\rB t-\sum_{m=1}^\infty\beta_m\sin (m\omega t) \ ,\quad
\beta_m=f_m/m\omega\,.
\end{equation}
This yields 
\begin{eqnarray}
\chi_t&=&g\int_0^t \re^{-\ri \eta_\tau}\,\rd \tau 
=g\int_0^t \re^{-\ri \omega_B\tau+\ri \sum_{m=1}^\infty\beta_m\sin (m\omega \tau)}\,\rd \tau\nonumber\\
&=&g \sum_{\nu =-\infty}^{+\infty}
J_\nu(\{\beta_m \})\int_0^t\re^{-\ri (\omega_\rB -\nu \omega)\tau}\,\rd \tau\
\end{eqnarray}
in terms of the infinite-variable Bessel functions \cite{Lore95,02tb2d} 
\begin{equation} 
\exp \Big(\ri \sum_{m=1}^\infty \beta_m\sin mu \Big) = \sum_{\nu =-\infty}^{+\infty}
J_\nu(\{\beta_m \})\,\re^{-\ri \nu u}\,.
\end{equation}
For resonant driving, $\omega_\rB=n\omega$, the final result is
\begin{equation}
\chi_t
=gJ_n (\{\beta_m \})t + 2g \sum_{\nu\ne n} J_\nu (\{\beta_m \})
\,\frac{1}{\omega_\nu}\,\re^{-\ri\,\omega_\nu t/2}\sin \frac{\omega_\nu t}{2}\,,
\label{inf-besselint}
\end{equation}
where we have set $\omega_\nu=\omega_\rB-\nu \omega$.  
In the non-resonant case the first term is absent and the sum extends over all
integers. 
Specializing again to mono- or bichromatic driving this result reduces to the
ones derived in sect.~\ref{s-bicromatic}.

For resonant driving, $\omega_\rB=n\omega$, we have $\omega_\nu =(n-\nu)\omega$
and the oscillating part in (\ref{inf-besselint}) vanishes at times which are
integer multiples of the driving period $T$.
The average transport velocity is then given by
\begin{equation}
v_{\rm trans} =|\gamma|=\frac{2|\chi_T|}{T}=\big|g J_n (\{\beta_m \})\big|\,.
\label{inf-besselint-v}
\end{equation}
Dynamical localization effects will be observed if the system parameters 
are tuned to a zero of the infinite-order Bessel function $J_n (\{\beta_m \})$. 

The closed form result for the transport velocity in (\ref{inf-besselint-v}) is
a generalization of the formula (\ref{vtrans-bicro}) derived for bichromatic driving.
For the very special case of a flipped rectangular field, the closed form
solution presented in section \ref{ss-flip} is, of course, superior to the
general solution.
\section{Concluding remarks}
We have shown, that
the dynamics of a quantum particle in a one-dimensional periodic
potential (lattice) under the action of a static and time-periodic field shows
quite intricate transport and localization phenomena, even in the relatively simple
case of bichromatic driving which depends sensitively
on the communicability relations between the two excitation frequencies and the 
Bloch frequency. The theoretical analysis based on  
a nearest-neighbor tight-binding model 
allows a convenient closed form
description of the transport properties in terms of generalized Bessel functions. 
In particular, conditions for the system parameters leading to localization are derived.

The case of polychromatic driving is also discussed, in
particular for flipped static fields, i.e.~rectangular pulses, which can support an
almost dispersionless transport with a velocity independent of the field amplitude.  

Such fields, which can be quite easily realized experimentally, offer interesting
possibilities for the control of quantum transport and the manipulation of wave packets
as for instance cold atoms in optical lattices. 

Finally it should be noted that it offers also some advantage to modify the
spatial period of the lattice. For example, a double periodic lattice, a superposition
of two periodic structures with period  $d$ and $2d$ has been studied recently
\cite{06bloch_zener,06bloch_manip}. It allows controlled Landau-Zener tunneling
which can also be employed for a manipulation of matter waves.

\section*{Acknowledgments}
Support from the Deutsche Forschungsgemeinschaft
via the Graduiertenkolleg  ``Nichtlineare Optik und Ultrakurzzeitphysik''
as well as from the ``Studienstiftung des deutschen Volkes''
is gratefully acknowledged.

\end{document}